\documentclass[iop]{emulateapj}

\usepackage{epstopdf}

\newcommand{\psizone}{1.15}

\slugcomment{Accepted by ApJ Letters}

\shorttitle{Mass-Dependence of Spin Evolution}
\shortauthors{Matt et al.}

\begin{document}

\title{The Mass-Dependence of Angular Momentum Evolution in Sun-Like Stars}

\author{Sean P. Matt$^{1}$,
        A. Sacha Brun$^{2}$, 
        Isabelle Baraffe$^{1,3}$, 
       J\'er\^ome Bouvier$^{4,5}$, and
        Gilles Chabrier$^{3,1}$}

\affil{$^1$University of Exeter, Department of Physics \&
   Astronomy, Physics Bldg., Stocker Road, Exeter, EX4 4QL,
   UK; s.matt@exeter.ac.uk}

\affil{$^2$Laboratoire AIM Paris-Saclay, CEA/Irfu Universit\'e
 Paris-Diderot CNRS/INSU, 91191 Gif-sur-Yvette, 
France}

\affil{$^3$\'Ecole Normale Sup\'erieure de Lyon, CRAL, 69364 Lyon Cedex 07, France}

\affil{$^4$Universit\'e de Grenoble Alpes, IPAG, F-38000 Grenoble, France}

\affil{$^5$CNRS, IPAG, F-38000 Grenoble, France}

\begin{abstract}

  To better understand the observed distributions of rotation rate and
  magnetic activity of sun-like and low-mass stars, we derive a
  physically motivated scaling for the dependence of the stellar-wind
  torque on Rossby number.  The torque also contains an
  empirically-derived scaling with stellar mass (and radius), which
  provides new insight into the mass-dependence of stellar magnetic
  and wind properties.  We demonstrate that this new formulation
  explains why the lowest mass stars are observed to maintain rapid
  rotation for much longer than solar-mass stars, and simultaneously,
  why older populations exhibit a sequence of slowly rotating stars,
  in which the low-mass stars rotate more slowly than solar-mass
  stars.  The model also reproduces some previously unexplained
  features in the period-mass diagram for the {\it Kepler} field,
  notably: the particular shape of the ``upper envelope'' of the
  distribution, suggesting that $\sim95$\% of {\it Kepler} field stars
  with measured rotation periods are younger than $\sim$4 Gyr; and the
  shape of the ``lower envelope,'' corresponding to the location where
  stars transition between magnetically saturated and unsaturated
  regimes.


\end{abstract}


\section{Introduction} \label{sec_introduction}

This Letter presents a formulation for the global angular momentum
loss of sun-like stars, defined here as stars with less than $\sim 1.3
M_\odot$, which have outer convective envelopes and are magnetically
active.  The goal is to develop a comprehensive physical model for the
evolution of stellar angular momentum that (a) explains both the
age-dependence and mass-dependence of observed stellar spin rate
distributions and (b) is fully consistent with our current best
understanding of stellar wind dynamics, magnetic properties, and
mass-loss rates.

The work is both motivated and enabled by large samples of stellar
rotation periods, now existing for several clusters, spanning an age
range of $\sim 10^{6-9}$~yr
\citep[][]{Irwin:2009p1363,Bouvier:2013p6008}.  When plotted in
period-color (or period-mass) diagrams, the distributions exhibit a
complex but apparently coherent evolution with cluster age
\citep{Barnes:2003p1513}.  This evolution includes a relatively smooth
dependence on stellar mass, from $\sim$1.3~$M_\odot$ down to the
substellar limit.  In general, during the first several hundred Myr,
lower mass stars take longer to spin down than higher mass stars.
Second, and somewhat paradoxically, after $\sim100$~Myr, the slowest
rotators begin to converge toward a narrow ``sequence'' in which the
lower-mass stars rotate more slowly than higher-mass stars.  This
behavior, particularly of the slowly-rotating sequence, gave birth to
gyrochronology
\citep{Barnes:2003p1513,Soderblom:1983p4391,Skumanich:1972p4350}, the
idea that stellar ages may be inferred solely from rotation period and
mass.  Gyrochronology will become increasingly important for recent
and future datasets (e.g., from exoplanet transit searches) that
provide rotation period measurements of large samples of stars with
unknown ages.  The best current example is the measurement of 34,000
rotation periods in the {\it Kepler} mission field of view, by
\citet{Mcquillan:2014p5988}.

The present model builds upon many previous works, including:
theoretical developments of how magnetized stellar winds remove
angular momentum
\citep{Schatzman:1962p4464,Mestel:1968p4497,Weber:1967p3752,Kawaler:1988p1012,Matt:2012p4894};
models for the evolution of stellar spin rate in time
\citep[e.g.,][]{MacGregor:1991p4316,denissenkovea10,Scholz:2011p4064,Reiners:2012p4157,Gallet:2013p5075,vanSaders:2013p5163,Brown:2014p6125};
and gyrochronology relations
\citep{Barnes:2003p1513,Barnes:2010p1157,Mamajek:2008p4506,Meibom:2009p1364}.

Much of the difficulty in predicting stellar wind torques arises from
the uncertainty (both observational and theoretical) in our knowledge
of the magnetic and stellar-wind properties of stars.  Despite
significant progress in measurements of mass-loss rates of sun-like
stars \citep{Wood:2005p1261}, theoretical predictions of wind
properties \citep{Suzuki:2012p5032,Cranmer:2011p3830}, mapping of
surface magnetic fields \citep{donatilandstreet09}, and dynamo models
\citep[][]{Miesch:2009p5057,Brun:2014p6136}, we are still working to
understand how these properties depend upon stellar mass, rotation
rate, and time.  Observations of various indicators of magnetic
activity
\citep{Noyes:1984p6120,Reiners:2009p538,Pizzolato:2003p3630,Mamajek:2008p4506,Wright:2011p48,Vidotto:2014p6054},
as well as theoretical models for magnetic field generation
\citep{Durney:1978p6126,Noyes:1984p6128,Baliunas:1996p6124,Jouve:2010p1250},
suggest that a key parameter for stellar magnetism is the Rossby
number,
\begin{eqnarray}
\label{eq_rossby}
Ro\equiv (\Omega_*\tau_{cz})^{-1},
\end{eqnarray}
where $\Omega_*$ is the angular rotation rate of the star, and
$\tau_{cz}$ is the convective turnover timescale, characterized by the
size of a convective region divided by the convective velocity.  For
slowly rotating stars, magnetic properties appear to correlate
strongly with $Ro$.  Below a critical value of the Rossby number,
$Ro_{\rm sat}$, various magnetic activity indicators appear to
``saturate,'' in a sense that they have an approximately constant,
maximal value (independent of $Ro$).  The value at which the
saturated/unsaturated transition occurs can be specified by a constant
\begin{eqnarray}
\label{eq_chi}
\chi\equiv {Ro_{\odot}\over Ro_{\rm sat}}\equiv 
  {\Omega_{\rm sat}\tau_{cz}\over \Omega_\odot\tau_{cz\odot}},
%
\end{eqnarray}
where ``$\odot$'' refers to solar values.  Saturation occurs for
$Ro\le Ro_{\odot}/\chi$, and $\chi$ defines the critical rotation
rate, $\Omega_{\rm sat}$ (or period $P_{\rm sat}\equiv
2\pi/\Omega_{\rm sat}$), for any star with known
$\tau_{cz}/\tau_{cz\odot}$.  The various studies cited above suggest
that $\chi$ lies in the approximate range of 10--15.

The model presented here reproduces some previously unexplained
features in period-mass diagrams and also places constraints on the
scaling of magnetic activity with Rossby number and stellar mass.

\section{Stellar Wind Torque Model}

\subsection{General Formulation} \label{sec_generaltorque}

Models of stellar wind dynamics
\citep{Kawaler:1988p1012,Matt:2012p4894} show that the torque on the
star can be written generically,
\begin{eqnarray}
\label{eq_torque}
%
%
%
T=T_\odot\left({M_*\over M_\odot}\right)^{-m}\left({R_*\over R_\odot}\right)^{5m+2}\nonumber \\
    \times \left({B_*\over B_\odot}\right)^{4m}\left({\dot M_w\over \dot M_\odot}\right)^{1-2m}\left({\Omega_*\over \Omega_\odot}\right),
\end{eqnarray}
where $M_*$ and $R_*$ are the stellar mass and radius, $B_*$ the
magnetic field strength on the stellar surface, and $\dot M_w$ the
global mass outflow rate.  The exponent factor $m$ is determined
primarily by the magnetic field geometry and wind acceleration profile
\citep{Reville:2014p6182} and likely falls in the range $m=$0.20--0.25
\citep{Washimi:1993p3825,mattpudritz08II,uddoula3ea09,Pinto:2011p4062,Matt:2012p4894}.
 
Given the uncertainties in both $B_*$ and $\dot M_w$, we adopt a
generic, combined relationship, based upon the rotation-activity
phenomenology discussed in section \ref{sec_introduction},
\begin{eqnarray}
\label{eq_rotact}
%
%
%
%
\left({B_*\over B_\odot}\right)^{4m}\left({\dot M_w\over \dot M_\odot}\right)^{1-2m}=Q\left({Ro_{\odot} \over Ro}\right)^p
\;\;{\rm (unsaturated)}, \\
\label{eq_rotactsat}
\left({B_*\over B_\odot}\right)^{4m}\left({\dot M_w\over \dot M_\odot}\right)^{1-2m}=Q\chi^p\;\;{\rm (saturated)},
%
\end{eqnarray}
which inherits the degeneracy between $B_*$ and $\dot M_w$ from
equation (\ref{eq_torque}).  The exponent $p$ encapsulates the
dependence of this combined activity factor on Rossby number.  The
generic scale-factor $Q$ has a yet-unknown dependence on stellar
parameters, which is determined empirically in section
\ref{sec_obstorque}.

\begin{deluxetable}{ccl}
\tablewidth{0pt}
\tablecaption{Adopted Parameter Values \label{tab_adopted}}
\tablehead{
\colhead{Symbol} &
\colhead{Adopted Value} &
\colhead{Description}
}


\startdata  

$\chi$ & 10 & Inverse critical Rossby number for \\ && magnetic saturation (solar
units) \\
$p$ & 2 & Rotation-activity scaling, eq.\ (\ref{eq_rotact}) \\
$M_\odot$ & $1.99 \times 10^{33}$ g & Solar mass \\
$R_\odot$ & $6.96 \times 10^{10}$ cm & Solar radius \\
$\Omega_\odot$ & $2.6 \times 10^{-6}$ Hz & Solar (solid body) angular rot. rate \\
$I_\odot$ & $1.05 \times 10^{54}$ g cm$^2$ & Solar moment of inertia \\
$t_\odot$ & $4.55 \times 10^{9}$ yr & Solar age \\
$\tau_{cz \odot}$ & 12.9 d & normalization for conv.\ turnover time 

\enddata

\end{deluxetable}

A combination of equations (\ref{eq_rossby})--(\ref{eq_rotactsat})
results in a bifurcated equation for the stellar wind torque,
\begin{eqnarray}
\label{eq_tunsat}
T=-T_0\left({\tau_{cz}\over \tau_{cz\odot}}\right)^{p}\left({\Omega_*\over \Omega_\odot}\right)^{p+1}\;\;\;\;\;\;{\rm  
     (unsaturated)}, \\
\label{eq_tsat}
%
T=-T_0\chi^{p}\left({\Omega_*\over \Omega_\odot}\right)\;\;\;\;\;\;{\rm (saturated)},
\end{eqnarray} 
where $T_0=T_0(T_\odot,M_*,R_*,Q, m$) does not depend upon the spin
rate or $\tau_{cz}$.  For the remainder of this work, we adopt
$\chi=10$, consistent with rotation-activity relationships and within
the range used in spin-evolution models cited in section
\ref{sec_introduction}.  We also adopt $p=2$, which gives the
unsaturated spin-scaling ($T\propto\Omega_*^3$) most commonly used in
the literature.  Table \ref{tab_adopted} lists the value of all
adopted parameters in the present work.

\subsection{Observationally Inferred Torque-Scaling} \label{sec_obstorque}

It is clear from the derivation above that $T_0$ should have a complex
dependence on stellar parameters, depending on $m$ and $Q$.  Given the
uncertainty associated with these quantities, we used the observed
stellar spin rates to infer a dependence of $T_0$ on stellar mass.  We
tested various scalings for $T_0$ and settled on one that is a
compromise between physical motivation and simplicity.  Specifically,
we adopt
\begin{eqnarray}
\label{eq_t0obs}
T_0=9.5\times 10^{30}\;{\rm erg}\;\;\left({R_*\over R_\odot}\right)^{3.1}\left({M_*\over M_\odot}\right)^{0.5}.
\end{eqnarray}
For the empirically derived scaling of equation (\ref{eq_t0obs}) to be
consistent with equations (\ref{eq_torque})--(\ref{eq_tsat}), the
general formulation requires that $T_\odot=9.5\times 10^{30}$ erg
and
\begin{eqnarray}
\label{eq_qobs}
Q=\left({R_*\over R_\odot}\right)^{3.1-(5m+2)}\left({M_*\over M_\odot}\right)^{0.5+m}.
\end{eqnarray}

\subsection{Analysis of Spin-Down in Time} \label{sec_analytic}

\begin{figure}
\epsscale{\psizone}
\plotone{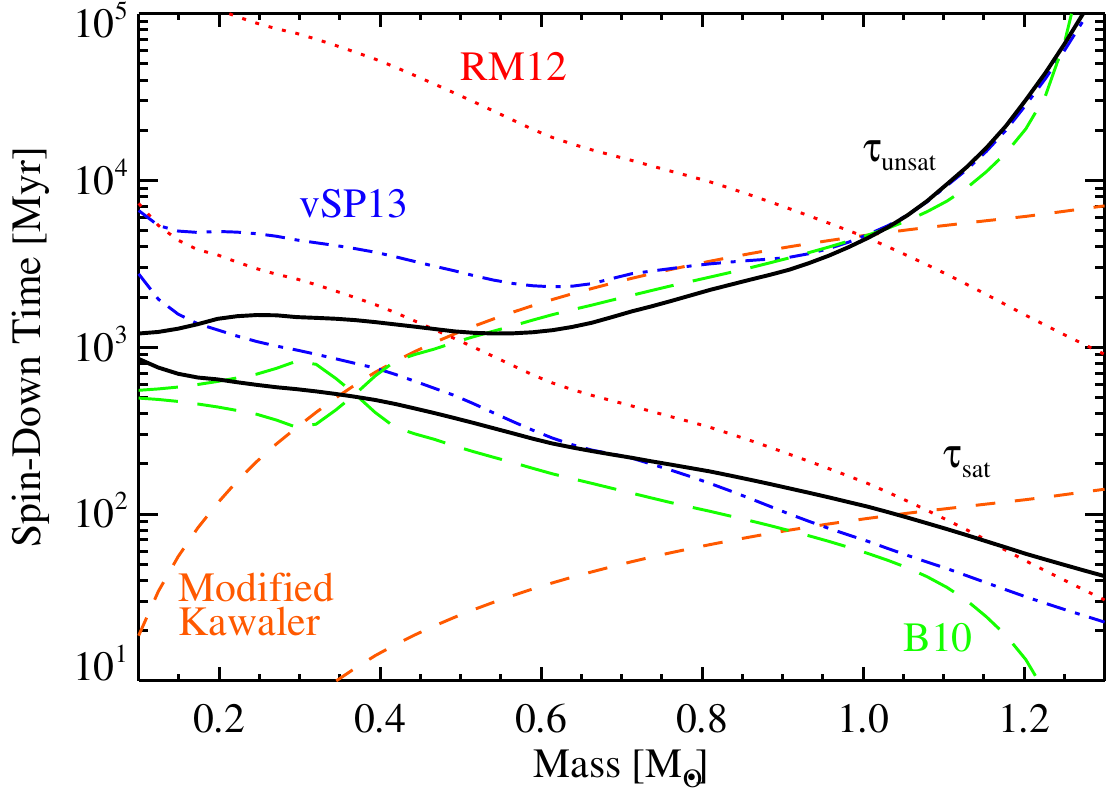}
\caption{Spin-down time in the saturated ({\it lower solid line};
  fast-rotation) and unsaturated ({\it upper solid line};
  slow-rotation) regimes, versus stellar mass.  The overall slope of
  $\tau$ versus $M_*$ has an opposite sign in each regime.  This
  slope-change explains why lower-mass stars remain rapid rotators for
  longer than higher-mass stars and, simultaneously, why
  slowly-rotating stars form a sequence in which the lower-mass stars
  rotate more slowly than higher-mass stars.  The broken lines show
  the two spin-down times for other models in the literature: {\it
    dashed}---modfied Kawaler formulation
  \citep[e.g.,][]{vanSaders:2013p5163}; {\it
    long-dashed}---\citet[][B10]{Barnes:2010p1157}; {\it
    dotted}---\citet[][RM12]{Reiners:2012p4157}; {\it
    dot-dashed}---\citet[][vSP13]{vanSaders:2013p5163}}
\label{fig_tausd}
\end{figure}

Using the torque defined by equations
(\ref{eq_tunsat})--(\ref{eq_t0obs}), we can now solve an angular
momentum equation to obtain the spin rate of any star as a function of
time, $t$.  Under the simplifying assumptions of solid-body rotation
and that the stellar moment of inertia, $I_*$, is constant in time
(approximately true for main-sequence stars), there are analytic
solutions given by
\begin{eqnarray}
\label{eq_sdsat}
\Omega_*=\Omega_i e^{-t/\tau_{\rm sat}}\;\;\;\;\;\;{\rm (saturated)}, \\
\label{eq_omegaconverged}
\lim_{\Omega_*\ll \Omega_{\rm sat}}\left({\Omega_*\over \Omega_\odot}\right) \rightarrow 
\left({\tau_{\rm unsat}\over t}\right)^{1\over p}\;\;\;\;\;\;{\rm (unsaturated)}, 
\end{eqnarray} 
where $\Omega_i$ is the ``initial'' spin rate, corresponding in
practice to some very young age ($t\ll \tau_{\rm sat}$), and two
spin-down timescales are defined as
\begin{eqnarray}
\label{eq_tausat}
\tau_{\rm sat}\equiv {I_*\Omega_\odot\over T_0\chi^{p}}\\
\label{eq_tauunsat}
\tau_{\rm unsat}\equiv {I_*\Omega_\odot\over T_0 p}\left(\tau_{cz\odot}\over \tau_{cz}\right)^p.
\end{eqnarray}
Equation (\ref{eq_omegaconverged}) predicts the spin rate only in the
asymptotic limit of $\Omega_*\ll \Omega_{\rm sat}$.  Stars generally
begin their lives with rotation rates in the saturated regime.
Equation (\ref{eq_sdsat}) then applies until a time when the spin rate
decreases to the critical spin rate, $\Omega_{\rm sat}$, after which
all spin rates asymptotically converge and approach equation
(\ref{eq_omegaconverged}).  This converged spin rate is independent of
the initial value, $\Omega_i$, and decreases as a simple power-law in
time, reproducing the \citet{Skumanich:1972p4350} relationship for
$p=2$.

To illustrate the effect of the torque in each regime,
Figure~\ref{fig_tausd} shows the spin-down times (eqs.\
[\ref{eq_tausat}] and [\ref{eq_tauunsat}]), as a function of stellar
mass.  For the Figure, we use values of $I_*$ from stellar models of
\citet{Baraffe:1998p5181} at an age of 2 Gyr, and compute $\tau_{cz}$
using the model effective temperatures with equation (36) of
\citet{Cranmer:2011p3830}.  The saturated spin-down time ({\it lower
  line}) represents the $e$-folding time of the spin rate, since the
spin-down is approximately exponential (eq.\ [\ref{eq_sdsat}]).  Once
stars are in the unsaturated regime, $\tau_{\rm unsat}$ ({\it upper
  line}) corresponds to the age at which the converged spin rate
equals the solar rate, $\Omega_\odot$; $\tau_{\rm unsat}$ also
predicts the mass-dependence of the converged spin rates, at any age
(according to eq.\ [\ref{eq_omegaconverged}]).

Since $\chi$ and $p$ are constants, the difference in the
mass-dependence of $\tau_{\rm sat}$ and $\tau_{\rm unsat}$ is due {\it
  entirely} to the factor of $\tau_{cz}^p$ (appearing only in
$\tau_{\rm unsat}$).  This difference is enough to reverse the sense
of the mass-dependence in the two regimes: Higher-mass stars spin-down
the most quickly in the saturated regime, but in the unsaturated
regime, lower-mass stars spin-down the most quickly.

Figure \ref{fig_tausd} also shows the equivalent spin-down times for
some models in the literature, with $\tau_{\rm unsat}$ normalized to
the sun.  Of these, the vSP13 and B10 models are most similar to the
present model, in both saturated and unsaturated regimes.  However,
all models differ by more than a factor of 2, in some mass range.  All
models therefore predict significantly different spin-down behavior,
and the present model has been tuned (via eq.\ [\ref{eq_t0obs}]) to
best reproduce the observed phenomenology presented in section
\ref{sec_synthetic}.  The key strength of the present model is its
formulation, which connects the observed spin evolution to the scaling
of magnetic field strengths and mass loss rates.

\section{Evolution of a Synthetic Cluster} \label{sec_synthetic}

\begin{figure*}
\epsscale{\psizone}
\plottwo{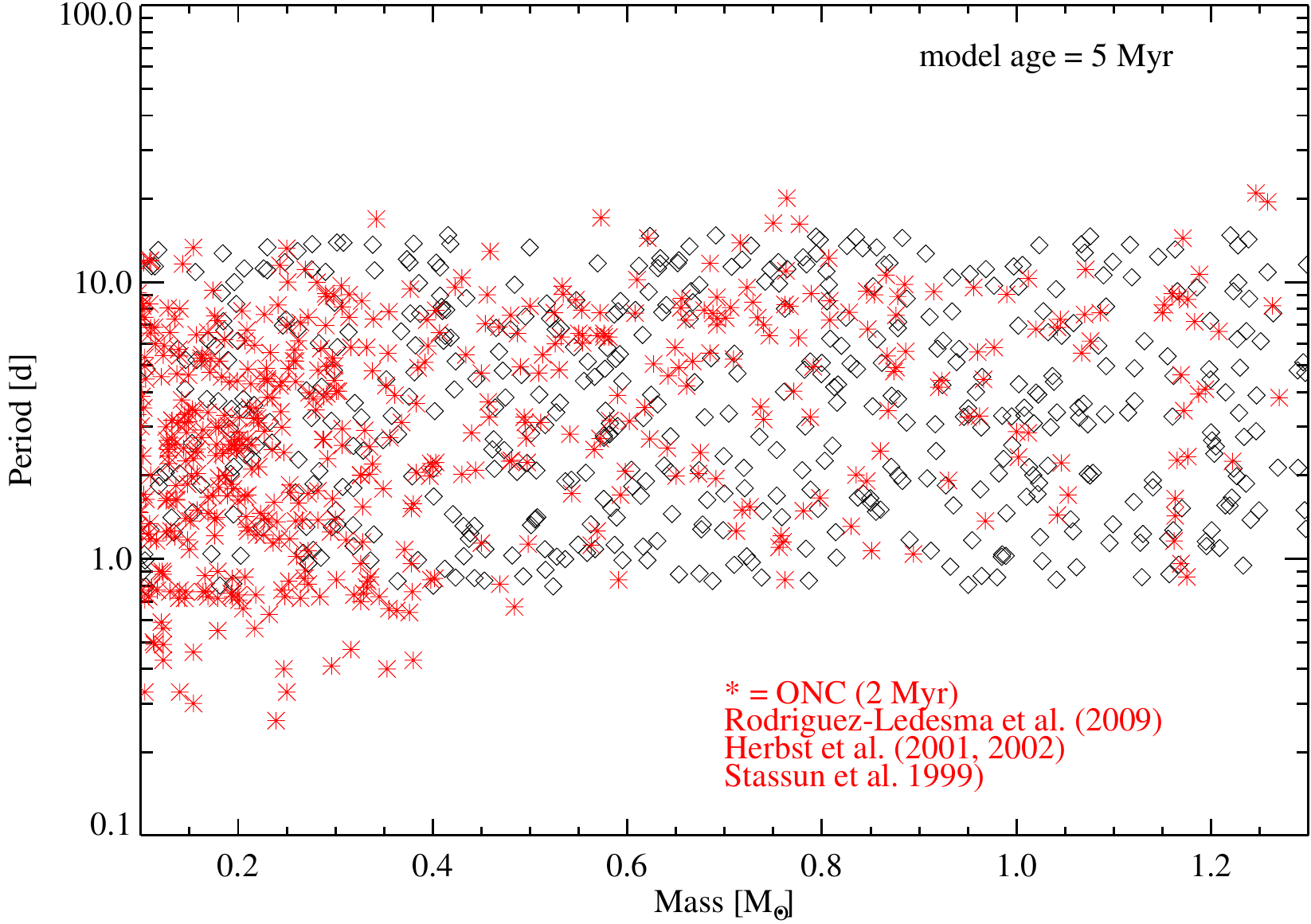}{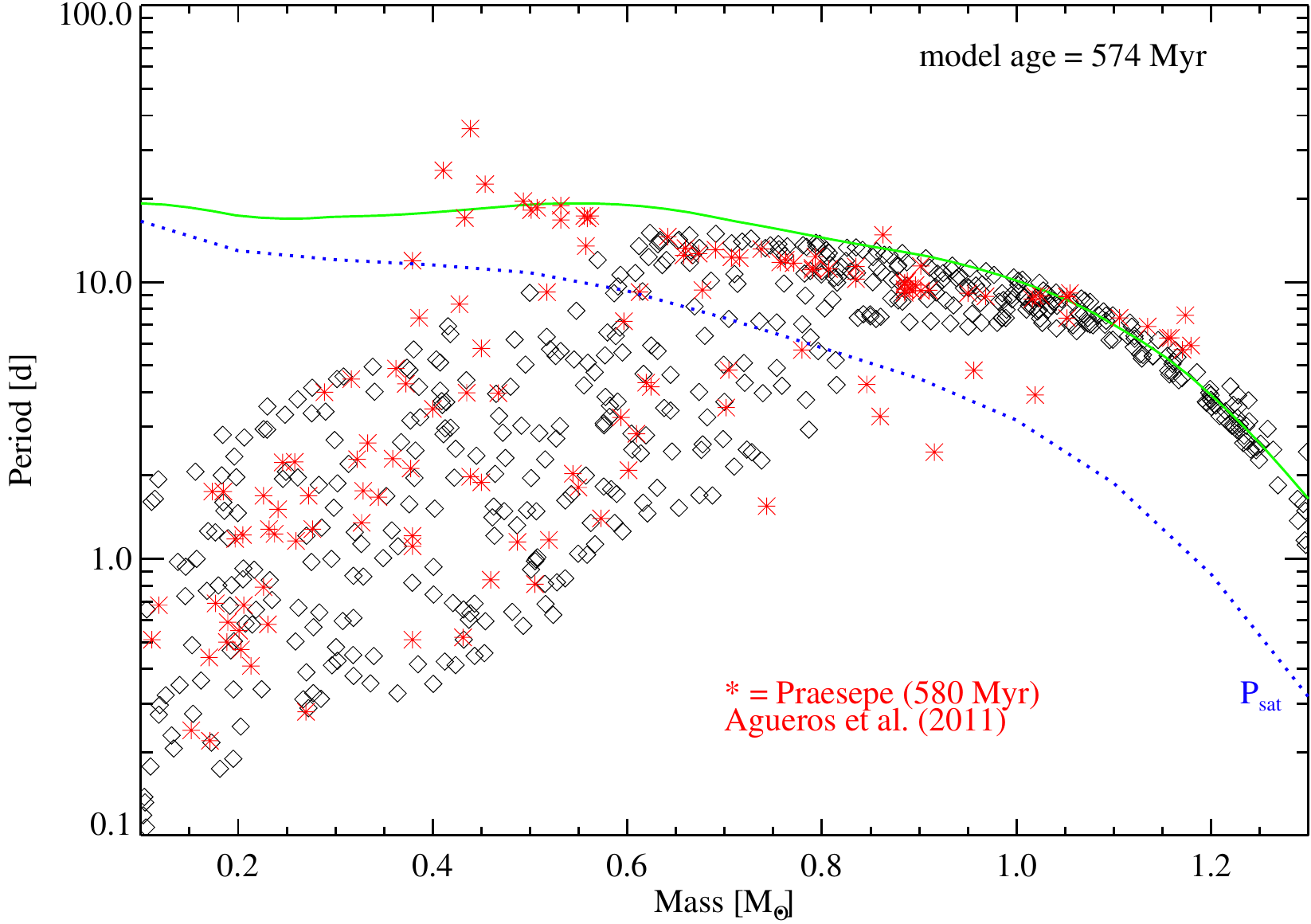}
\caption{Observed rotation periods (red stars) from the ONC (left
 panel) and Praesepe (right panel), compared to our synthetic cluster
 stars (black diamonds).  The left panel shows the synthetic initial
 conditions, chosen to approximate the observed range, but not the
 detailed distribution.  The right panel shows the synthetic cluster,
 evolved to a similar age as Praesepe (as indicated).  For reference,
 the green solid line shows the theoretical asymptotic spin rate of
 equation (\ref{eq_omegaconverged}), and the blue dotted line
 delimits magnetically saturated and unsaturated stars.  The model
 explains both the existence of rapidly rotating, low-mass stars, as
 well as the general mass-dependence of the slow-rotator sequence.}
\label{fig_clusterev}
\end{figure*}

\subsection{Initial Conditions}

To compare with observations, we computed the evolution of stellar
spin rates for a synthetic cluster of 500 stars.  We started the
evolution from an age of 5 Myr, in order to avoid the earliest phases,
where the spin distributions are poorly understood (and likely due to
processes not included here).  The cluster initially has a random and
uniform distribution in stellar mass (in the range 0.1-1.3 $M_\odot$)
and in the logarithm of rotation period (in the range 0.8-15 days).
The left panel of Figure \ref{fig_clusterev} shows this initial
distribution, compared with the $\sim2$ Myr-old cluster ONC
\citep[data
from][]{stassunea99,Herbst:2001p6097,herbstea02,RodriguezLedesma:2009p5348}.
The Figure demonstrates that the initial conditions approximate the
general range of rotation periods observed in young clusters, with no
attempt to fit or explain the detailed distribution.

\subsection{Spin Evolution} \label{sec_evolution}

Starting from the initial condition, we solved the angular momentum
equation
\begin{eqnarray}
\label{eq_amom}
{d\Omega_*\over dt}={T\over I_*}-{\Omega_*\over I_*}{d I_*\over dt},
\end{eqnarray}
for each star, using a forward-timestepping Euler method, and assuming
solid-body rotation.  The torque was specified by equations
(\ref{eq_tunsat})--(\ref{eq_t0obs}) (and eq.\ [\ref{eq_chi}]
determining the saturated/unsaturated transition) and values in Table
\ref{tab_adopted}.  At each timestep, we interpolated the stellar
parameters $R_*$, $I_*$, and $dI_*/dt$ from a grid of pre-computed
(non-rotating) stellar evolution tracks of \citet{Baraffe:1998p5181}
and computed $\tau_{cz}$ from the prescription of
\citet{Cranmer:2011p3830}.

The evolution proceeds as follows.  During the first several tens of
Myr, all stars are contracting and spin up by a factor of 5--10, as
they approximately conserve angular momentum (the torques are
negligible on this timescale).  When the stars reach the main
sequence, their structure stabilizes and they begin their spin-down.
Once $\Omega_* < \Omega_{\rm sat}$, their spin rates rapidly converge
toward the asymptotic spin rate predicted by equation
(\ref{eq_omegaconverged}).  This evolution, and the formation of a
converged, slow-rotator sequence, happens first for the highest mass
stars and proceeds in a continuous manner toward lower masses.  Figure
\ref{fig_clusterev} (right panel) and Figure \ref{fig_mea14_kepfield}
show the synthetic cluster after it has evolved to ages between
500~Myr and 4~Gyr.

\subsection{Comparison with Observations}

\subsubsection{Praesepe Cluster} \label{sec_praesepe}

The right panel of Figure \ref{fig_clusterev} compares the rotation
periods in the $\sim$580~Myr-old Praesepe cluster \citep[observed
by][]{Agueros:2011p6055} to the synthetic cluster, at a similar age.
Two key observed features are reproduced by the synthetic cluster.
First, there is a population of rapid rotators, exhibiting a wide
range of rotation rates and a trend such that the lowest mass stars
are, on average, more rapidly rotating than higher mass stars.  In the
models, the wide range is a consequence of the initial distribution,
but the trend with mass is due to the fact that lower mass stars take
longer to spin down, in the saturated regime (see Fig.\
\ref{fig_tausd}).

The second feature reproduced by the models is the population of stars
that have converged onto a relatively narrow sequence (following an
approximate upper limit in period).  In the models, the existence of a
converged sequence is due to the stars entering the unsaturated
regime, where the torque depends strongly upon rotation rate.  The
trend of rotation rate with mass is due to the fact that lower mass
stars generally spin down quicker than higher mass stars, once in the
unsaturated regime (Fig.\ \ref{fig_tausd}).

A few observed features are not reproduced by the model.  The first is
a handful of stars rotating more rapidly than the synthetic cluster
stars (in the range 0.7--1 $M_\odot$), which suggests a modified
torque for these stars.  Second is the population of slow rotators (in
the range 0.35--0.6 $M_\odot$) that appear to extend the slow-rotator
sequence to lower masses than in the synthetic cluster.  This
discrepancy likely arises from a deviation from solid-body rotation
(which the model assumes).  Studies that included internal angular
momentum transport
\citep{MacGregor:1991p4316,Gallet:2013p5075,Charbonnel:2013p5068,Denissenkov:2010p4340}
indicate that internal differential rotation manifests as an increased
spin-down at early times, followed by a convergence toward the
solid-body solution at later times.  The predicted, asymptotic spin
rate (green line in Fig.\ \ref{fig_clusterev}) roughly traces the
observed sequence over its full mass range, giving support for the
mass-dependence of the torque, even though the solid-body
approximation does not capture all details.

\subsubsection{{\it Kepler} Field}  \label{sec_kepler}

\begin{figure*}
\epsscale{.9}
\plotone{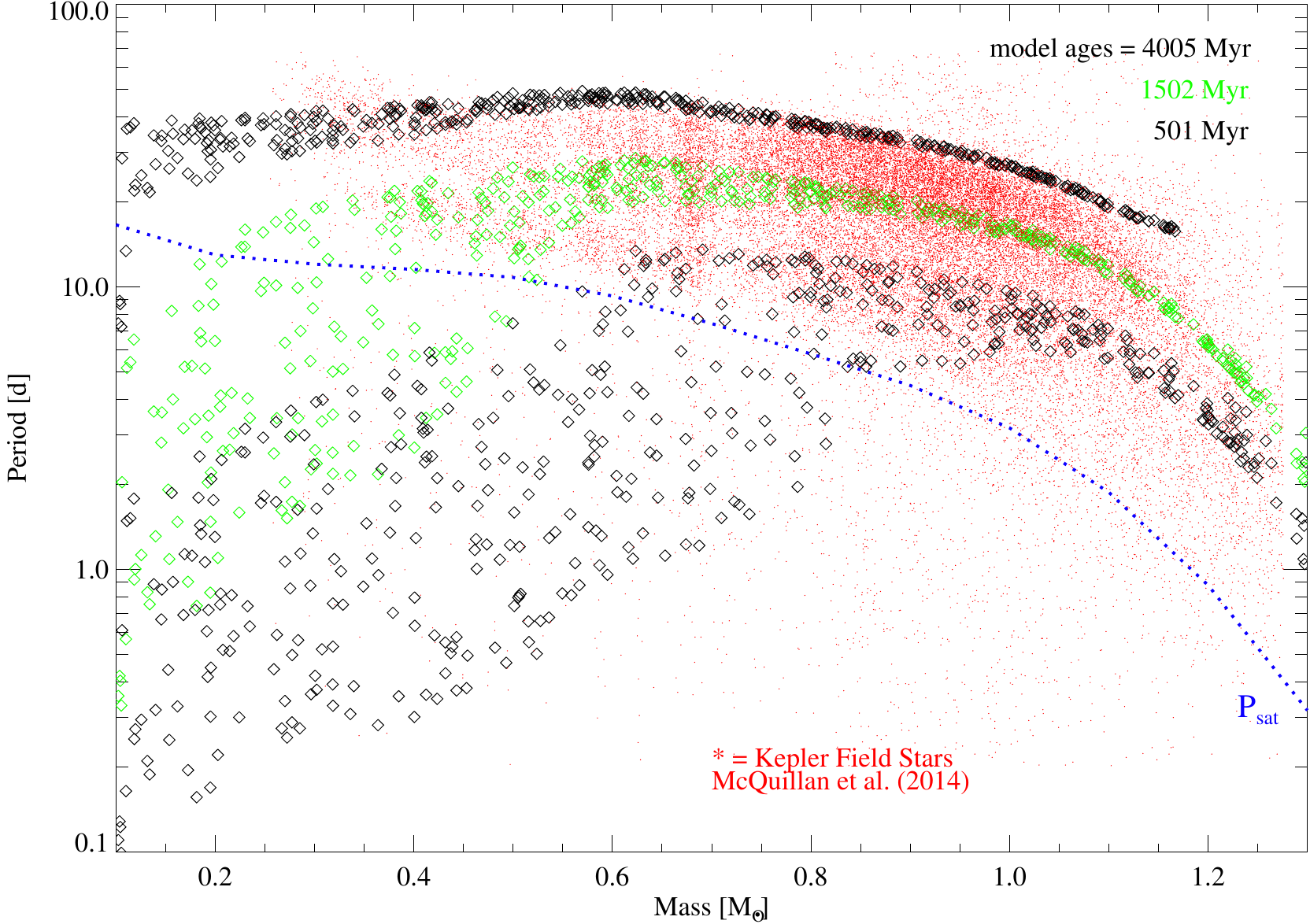}
\caption{Observed rotation periods in the {\it Kepler} field (red
  symbols), plotted over the syntetic cluster, shown at three
  different ages: 500 Myr (lower group of black diamonds), 1.5 Gyr
  (green diamonds), and 4.0 Gyr (upper group of black diamonds).  The
  blue dotted line shows the rotation period dividing the saturated
  and unsaturated regimes.  The coincidence of $P_{\rm sat}$ with the
  ``lower envelope'' of the {\it Kepler} stars, suggests that this
  feature is explained by the convergence of stellar spin rates,
  occurring after stars enter the unsaturated regime.  The coincidence
  of the oldest models with the observed ``upper envelope'' suggests
  that $\sim$95\% of the sample stars are younger than $\sim$4~Gyr.}
\label{fig_mea14_kepfield}
\end{figure*}

Figure \ref{fig_mea14_kepfield} compares the measured rotation periods
in the {\it Kepler} field \citep[][hereafter
MMA14]{Mcquillan:2014p5988} to the synthetic cluster, shown at three
different ages.  The figure only shows stars with measured rotation
periods, comprising 26\% of the total {\it Kepler} main-sequence
sample, and possessing a range of unknown ages.  Within the framework
of our model, we interpret some broad features of the observed spin
distribution in the {\it Kepler} field.

First, there is a well-defined ``upper envelope'' to the distribution
of observed rotation periods (corresponding approximately to the 95th
percentile of the distribution), which coincides with the 4~Gyr-old
synthetic cluster, for stars with $\ga 0.5 M_\odot$, including the
apparent ``dip'' or change in slope around 0.6 $M_\odot$.  This dip
has not been previously reproduced by any model.  The coincidence with
the model suggests that the existence and shape of the observed upper
envelope is real (rather than being due to observational bias) and
also corresponds to an age of $\sim$4~Gyr (also noted by MMA14).  At
masses below 0.5~$M_\odot$, the mismatch between the synthetic cluster
and observations indicates that the low-mass, unsaturated stars
require a stronger torque than the model predicts.

There is also a relatively sharp ``lower envelope'' in the observed
distribution of Figure \ref{fig_mea14_kepfield}, also noted by MMA14,
most pronounced for stars with $\la 0.9 M_\odot$.  This lower envelope
has not been previously explained, but it corresponds remarkably well
to the location of the critical rotation period (blue dotted line),
which delineates the saturated and unsaturated regimes in our model.
As apparent in the right panel of Figure \ref{fig_clusterev}, the spin
rates of stars begin to converge after crossing this critical rotation
period.  Thus, in a distribution of stars with a range of ages, the
model predicts that the density of stars will increase at a rotation
period slightly larger than the critical period, as observed.  Recall
that the critical rotation period (eq.\ [\ref{eq_chi}]) is set by a
constant saturation level, $\chi$, and the mass-dependent convective
turnover timescale, $\tau_{cz}$.  Thus, the coincidence of $P_{\rm
  sat}$ with the lower envelope of the {\it Kepler} spin distribution
supports the modeled relationship between convection, magnetic
activity (including saturation), and spin-evolution.  Furthermore,
{\it independent of any model}, the lower envelope coincides precisely
with the slow-rotator sequence observed in the youngest clusters in
which this feature appears \citep[those with ages of $\sim$100 Myr,
not shown;][]{Bouvier:2013p6008}.  This comparison with young
clusters, as well as with the present model, suggests that the {\it
  Kepler} field has a substantial population of stars with ages less
than $\sim$500 Myr (also noted by MMA14).

\section{Discussion and Conclusions} \label{sec_discussion}

The model presented here builds upon the ideas and successes of many
previous works (cited in \S \ref{sec_introduction}), notably in the
explanation for a saturation of the torque at high spin rates and a
Skumanich-style spin-down at later times.  However, the present model
provides a new formulation that reproduces some previously unexplained
phenomena, particularly related to the mass-dependence of observed
features in Figures \ref{fig_clusterev} and \ref{fig_mea14_kepfield}.

A number of observed phenomena that are not reproduced by the model
will require further improvements, for example: the model does not
well-produce the {\it Kepler} field slow rotators for masses below 0.5
$M_\odot$, which suggests (for example) that the adopted values of
$\tau_{cz}$ may not be appropriate for these stars; the overall
interpretation of the {\it Kepler} field star ages (\S
\ref{sec_kepler}) should be tested by population studies; a fraction
of stars (e.g., in Praesepe) appear to converge onto the unsaturated
sequence at an earlier time than the models, suggesting a deviation
from solid-body rotation; and the present model does not explain the
``initial'' conditions nor any of the more detailed structure present
in the spin distributions of young stars
\citep[see][]{Herbst:2001p6097,Henderson:2012p4069, Brown:2014p6125}.

Much of the success of the present model derives from the empirical
mass-scaling of the torque, given by equation (\ref{eq_t0obs}).  This
is not a unique solution, and the physics suggest a dependence on more
complex stellar properties than $M_*$ and $R_*$ (e.g., $\dot
M_w$ may depend on coronal Alfv\'en wave flux;
\citealp{Cranmer:2011p3830}).  However, for any other formulation to
work as well, the included physics must conspire to scale like
equation (\ref{eq_t0obs}).

Fitting the present model to observations provides constraints on the
physical parameters $\dot M_w$, $B_*$, $\chi$, $p$, and $m$, all of
which are connected to the physics and phenomenology of magnetic
properties and wind dynamics in sun-like stars.  For the parameters
adopted here and a dipolar magnetic field (i.e., $m=0.22$), the
model's torque could arise from the simple scalings\footnote{Formally,
  equations (\ref{eq_rotact}) and (\ref{eq_qobs}) define a family of
  solutions satisfying $B_*^{4m}\dot M_w^{1-2m}\propto
  R_*^{3.1-(5m+2)}M_*^{0.5+m}Ro^{-p}$.  We give one possibility here.}
$B_*\propto Ro^{-1}$ and $\dot M_w\propto M_*^{1.3}Ro^{-2}$.  These
scalings can be compared to models and observations and do not appear
unreasonable.  Thus, a key advantage of our formulation is that it
provides a basic framework for a self-consistent physical picture of
stellar evolution that includes the effects of magnetic activity, mass
loss, and rotation.



\acknowledgments

The EU's FP7 supported SPM and ASB, under grant \#207430 ``STARS2''
(http://www.stars2.eu), and IB, under grant \#320478 ``TOFU.''  JB and
ASB were supported by the grant ANR 2011 Blanc SIMI5-6 02001
``TOUPIES'' (http://ipag.osug.fr/Anr\_Toupies/).



\end{document}